**Superdipole Liquid Scenario for the Dielectric Primary Relaxation**

**in Supercooled Polar liquids**

Y.N. Huang[*], and C.J. Wang

*National Laboratory of Solid State Microstructures, Department of Physics, Nanjing University,*

*Nanjing* 210093*, P.R. China*

E. Riande

*Instituto de Ciencia y Tecnologia de Polimeros* (CSIC)*, 28006-Madrid, Spain*

**Abstract**. We propose a dynamic structure of coupled dynamic molecular strings for supercooled small polar molecule liquids and accordingly we obtain the Hamiltonian of the rotational degrees of freedom of the system. From the Hamiltonian, the strongly correlated supercooled polar liquid state is renormalized to a normal superdipole (SD) liquid state. This scenario describes the following main features of the primary or $\alpha$-relaxation dynamics in supercooled polar liquids: (1) the average relaxation time evolves from a high temperature Arrhenius to a low temperature non-Arrhenius or super-Arrhenius behavior; (2) the relaxation function crosses over from the high temperature exponential to low temperature non-exponential form; and (3) the temperature dependence of the relaxation strength shows non-Curie features. According to the present model, the crossover phenomena of the first two characteristics arise from the transition between the superdipole gas and the superdipole liquid. The model predictions are quantitatively compared with the experimental results of glycerol, a typical glass-former.

---
[*] To whom correspondence should be addressed. E-mail: ynhuang@netra.nju.edu.cn

# I. INTRODUCTION

Although the comprehensive understanding of the glass transition still remains a notorious unresolved problem in condensed matter physics and materials science [1-12], a fruitful and enlightening progress from both experimental and theoretical points of view has already been accomplished summarized in some reviews and special books [2,3,6,11].

The predominant issue regarding the glass transition is the description of the α-relaxation dynamics of the supercooled liquid state whose freezing leads to the thermodynamic glass transition [2-3,6,8, 10-12]. A great deal of experiments relevant to the α-relaxation show that [2,3,6,11]: (1) during vitrification the temperature dependence of the average relaxation time $\tau_\alpha$ evolves from high temperature Arrhenius to low temperature non-Arrhenius (super-Arrhenius) behavior which can successfully be described by the empirical Vogel-Fulcher equation [13]; (2) the relaxation function changes from high temperature exponential to low temperature non-exponential form described by the Kohlrausch-Williams-Watts empirical equation [14], or the Cole-Davidson equation [15] in the case of low molecular weight glass formers; and (3) the temperature dependence of the relaxation strength shows non-Curie features so it can approximately be fitted by the Curie-Weiss equation proposed by Chamberlin [16], named Currie-Weiss-Chamberlin equation hereafter. Moreover, it looks like some internal relations between the three characteristics indicated above exist [17]. All these features obviously deviate from the conventional relaxation theory of normal liquids, the well-known Debye theory [18], since according to this theory the relaxation time, the relaxation function and the relaxation strength should obey the Arrhenius relation, the exponential function and Curie law, respectively.

Some prevalent and enlightening theories related with the dynamic glass transition have already been reported. Among them, the Adam-Gibbs theory of cooperatively rearranging regions [19], the Cohen-Grest free-volume theory for percolation of solid clusters in a liquid matrix [20], the Ngais' coupling model [21], the Götzes' mode-coupling theory [22], the Kiveslsons' FLD model [23], the Chamberlin's mesoscopic mean-field theory [24], and Garrahan-Chandler coarse-grained microscopic model [12], etc. [25], stand out. However, it cannot be denied the existence of questionable and/or considerable points in these models or theories, which need to be clarified [2-3,11-17, 19-25].

The study of relaxation phenomena by broadband dielectric spectroscopy over a wide temperature range provides important insights into the mechanisms of the α-relaxation dynamics [11]. This technique is useful to investigate the relaxation phenomena of supercooled liquids, such as glycerol, a typical relatively simple glass-former [3] compared with polymers and other complex systems [26]. From a theoretical point of view, the conventional dielectric relaxation theory or Debye theory [18] provides a good start to study the relaxation behavior of the supercooled liquid state. In this paper, we model the three abnormalities of the dielectric relaxation of low molecular weight polar liquids mentioned above, and the organization of the paper is as follows. In Sec. II, we propose a dynamic structure of coupled dynamic molecular strings for supercooled polar liquids formed by small molecules, and based upon the structure we obtain a reduced Hamiltonian of the rotational degrees of freedom of the system. Sec. III contains solutions of the Hamiltonian and the results of the model. In Sec. IV we compare the theoretical predictions with experiments and discuss further the results of the model.

## II. MODEL

For a polar liquid of small rigid molecules, the Hamiltonian of the system can formally be expressed as $H(r_1, \varphi_1 \cdots r_k, \varphi_k \cdots)$, where $r_k$ and $\varphi_k$ are the translational and rotational coordinates of the $kth$ molecule, respectively [25]. In dielectric measurements, the external applied electric field directly couples to the rotational degrees of freedom of the molecules, but not to the translational movement, only induced by the rotational motion [11,18]. Specifically, the applied field induces orientational ordering of the molecules and this latter phenomenon further induces the translational ordering of the system, so that the latter ordering is a secondary effect of the former. In fact, an induced translational ordering is the well-known converse piezoelectric effect [18,27] or electrostriction effect [18,28]. In general, translational ordering is very small in the linear dielectric response regime of normal liquids, supercooled liquids and glasses. Consequently, as a first order approximation, the secondary induced translational movements of molecules can be neglected when we focus on the linear dielectric response of a glass-former like the Debye relaxation theory does [18].

In the study of the relaxation phenomena of normal liquids using the Debye theory, the induced secondary translational movements of molecules are omitted. Moreover, as an individual-particle mean-field approach, the complicated interaction between the rotational motions of a molecule and its neighbors is reduced to a double-well potential in which the dipole reorientates [18]. In this sense, the Hamiltonian of the system can be expressed as $H_0 = \frac{V_0}{2} \sum_{k=1}^{N_T} \{1 - \cos[2(\phi_k - \phi_k^0)]\}$, where $V_0 > 0$ is the activation barrier energy between the two wells, $N_T$ is the total number of molecules and $\phi_k$ $(0 \leq \phi_k < \pi)$ is the rotational angle of the *kth* dipole in the system. The isotropy of the system renders $\phi_k^0$ a uniform distributed quantity in the range $[0, \pi]$ [18]. The theory, which ignores the inter-dipole residual-rotational-correlation (RRC) of the individual-particle mean-field reduction, has achieved a great success in the description of the high temperature normal liquid state where molecules rotate so rapidly that approach the mean-field conditions quite well. This means that the RRC is small enough to be neglected. In supercooled liquids, where the rotational motions become slow, the RRC increases and therefore the dielectric relaxation dynamics of the supercooled liquid state is modified by the RRC.

The conventional individual-particle mean-field liquid theories, such as the cell model [29] and the hole model [30] as well as the significant structure theory [31], in which only the translational degrees of freedom are considered, present a successful description of the thermodynamics of the normal liquid state. However, an important recent finding, beyond the conventional liquid theories, is the existence of quasi-one-dimensional string-like cooperative molecular motions (molecular strings) widely observed in glass formers by well-designed experiments [32-33], analog simulations [34] and molecular dynamics simulations [35]. Additionally, there exists coupling between the strings [35]. From a thermodynamic point of view, the increase of viscosity with decreasing temperature leads to the suppression of Brownian motions and consequently, the decrease of the entropy of the system [20,30,36]. If the molecules move in a snakelike manner, i.e. one tagging after another, the interaction within the string will effectively reduce the internal energy of the system compared with the normal liquid state. On the other hand, snakelike motions confer these molecules the possibility of reaching more configurations, thus increasing the entropy of the system. Hence, it seems possible that translational snakelike movements could be another basic molecular motion manner in the supercooled state beyond the individual-molecule motions of the conventional mean-field liquid theories [29-31]. However, the physics behind this kind of motions is not clear yet.

In fact, Glotzer pertinently thinks that it is intriguing to consider the possibility that the strings may be the elementary cooperatively rearranging regions predicted by Adam and Gibbs [35]. As a general consideration, snakelike motions could be ascribed to the residual-translational-correlation between molecules after the individual-particle mean-field reduction of the conventional liquid theories [29-31].

According to molecular dynamics simulations, only a few percent of particles take the quasi-one-dimensional snakelike or string-like motions, the remaining particles intuitively behaving as located in a caged way forming domains [35]. It is worth noting that besides the fact that the fast particle criterion is more or less relative [35], its dynamics computation time is finite, e.g. it is only a few times larger than that of extrapolation of the high temperature Arrhenius relation [35,37]. So, another possible alternative scenario is to consider that the slow mobile molecules in the domains also move in a snakelike manner, though this scenario is not observed in the simulations because of the finite simulation time window [25,35]. In other words, we could think that all molecules move in a snakelike manner in the supercooled liquid state, i.e. a full string scenario such as the present model to be shown below. Furthermore, owing to the string length distribution and the coupling between the strings, there will exist relatively fast and slow mobile strings. For a full string scenario, the relaxation time of a string of 60 molecules is at 195 K about $10^4$ times larger than that of an adjacent string of 6 molecules for the typical glass former glycerol [3,11]. Therefore, it is expected that if the simulation time is the same as the relaxation time of the short string (Arrhenius-like relation in Sec. IV), the longer string will not relax in the time scale of the former. Moreover, due to the string length distribution as well as the fluctuation of the string distribution in space, it should be expected that some short or long strings, locally congregated in space due to the fluctuation, would couple forming spatial correlated regions (domains or clusters) in the system. These regions would show fractal morphology because of the quasi-one-dimensional characteristic of the congregating strings and their random stacking in space [25,32-38]. Therefore, the full string scenario does not seem to conflict with the simulations. The simulations in a large time scale and wide temperature window, doubtless an outstanding challenge ahead [35,37], would provide a criterion to assess the full string scenario and the picture of fast snakelike motion strings and slow mobile molecule domains.

The most basic problem of glass-liquid relaxation phenomena is the dynamic structure of glass-formers because it is the start of further calculations, and microscopic models, such as the mesoscopic mean-field theory, etc. [24], always face this problem. If we take the simulation results as a criterion, any microscopic model must contain the string-like or snakelike collective motions, otherwise it would be a more or less phenomenological model. In this kind of models, we would like to discuss two cases. One is the full string scenario of small molecule glass-formers, such as the present model to be shown below, in which the basic unit of the structure, i.e. the strings, is similar to the macromolecules of polymer glasses [36,39], so that the structure and its fabricating process are also similar. It is obvious that a unified picture based upon this scenario can be obtained for small molecules and polymer glasses. Another case is the picture containing strings of snakelike fast mobile molecules and domains of slow mobile molecules. Obviously, the structure fabrication of this picture is more complicated than that of the first one for we need to develop domains with certain structures, besides the molecular strings, and stack them appropriately in three-dimensional space. Moreover, the α-relaxation and the glass transition phenomena are similar, at least qualitatively, for both low molecular weight glasses and polymer glasses [11,26], and it is well known that these phenomena are closely related to the segmental motions in the latter materials. Therefore, the full string scenario seems to be a reasonable hypothesis.

As for the collective motion of dipole rotations arising from the inter-dipole residual-rotational-correlation (RRC) of the individual-particle mean-field reduction of the Debye theory, recent simulations also show the rotational string-like behavior of molecules [37]. Here we propose, besides the individual-dipole mean-field reorientations of the Debye theory, the following hypothesis: (1) the reorientation of all dipoles exhibits snakelike behavior and the spatial configurations of the orientational strings behave like a self-avoiding (i.e. the excluded volume effect) free rotational chain [36,39]; and (2) there is secondary coupling between strings. Physically, the Hamiltonian of the system related to the rotational degrees of freedom $H$ can be expressed as the sum of the zero-order Hamiltonian of the mean-field individual-dipole reorientation of the Debye theory $H_0$ [18], the first-order Hamiltonian of the orientational strings of dipoles $H_1$ and the second-order Hamiltonian of the coupling between the strings $H_2$, i.e. $H = H_0 + H_1 + H_2$.

In the temperature range of the supercooled liquid state [18] $e^{V_0/T} >> 1$ (here we use the system of units that sets the Boltzmann constant equal 1), most dipoles will be located in one of the double-wells of $H_0$, thus rendering possible the use $\sigma = \pm 1$ to denote the orientation states of the dipoles. Moreover, since the interaction between dipoles related to the rotational degrees of freedom is of short range order in structural glasses, only the nearest neighboring interactions need to be considered [24]. Then, the model Hamiltonian of the system related to the reorientation of dipoles can be written as,

$$H = H_0 + H_1 + H_2$$
$$H_0 = \frac{V_0}{2} \sum_{k=1}^{N_T} \left\{ 1 - \cos[2(\phi_k - \phi_k^0)] \right\}$$
$$H_1 = -V_1 \sum_m \sum_{k=1}^{n-1} \sigma_k^{mn} \sigma_{k+1}^{mn} \tag{1}$$
$$H_2 = \frac{V_2}{2} \sum_m \sum_{k=1}^{n} \sum_{m' \neq m, l}^{NN(k)} \sigma_k^{mn} \sigma_l^{m'n'} \cos \alpha_{kl}^{mm'}$$

where $H_1$ and $H_2$ describe the intra- and inter-string interactions, respectively. $V_1$ and $V_2$ are positive constants independent of temperature. The symbol $\sigma_k^{mn} = \pm 1$ denotes the orientation states of the $k$th dipole in a string labeled $m$ with molecular number $n$ in the system (called $n$-string hereafter). $NN(k)$ represents the nearest number of dipoles surrounding dipole $k$ that is determined by the average coordination number $z$, and $\alpha_{kl}^{mm'}$ is the angle between the dipole $k$ in the $n$-string and the dipole $l$ in the $n'$-string.

$V_1$ in $H_1$ only describes the connecting ability between two adjacent dipoles in the strings (Sec. III-2). Concerning the spatial configuration of the orientational strings, we need another parameter, the directional angle of the chain $\theta$ [36,39], to describe the self-avoiding free rotational chain behavior of the strings mentioned above. It will be shown latter that both the string length distribution (Sec. III-2) and the effective coupling between strings (Sec. III-3) are closely related to $\theta$. $V_2$ indicates the coupling strength between strings. One would expect that this coupling, the string length distribution and the distribution of strings in space will lead to the formation of spatial clusters of coupled strings (Sec. IV).

According to the hole model [30] and the significant-structure theory [31], there are quite a large number of molecular holes in liquids that remind the well-known free-volume theory [20,40]. On the other hand, for a string with a finite number of dipoles there are two end dipoles corresponding to the termination of the intra-string correlation. In moving a dipole away from an inner string, the two neighboring dipoles to the molecular hole in the string become non-correlated (or at least very small correlated) and the single string becomes two strings. Thus, the formation of strings may be closely related to the molecular holes in the supercooled liquid state, and owing to the random movements of the holes arising from thermal agitation, both the distribution of the strings in space and the string length distribution are dynamic quantities.

There are three issues relevant to the Hamiltonian (Eq.1): (i) intra-string correlation, (ii) string-length distribution, and (iii) inter-string correlation. In Sec. III we discuss each of them.

## III. MODEL SOLUTIONS AND RESULTS

### III-1. Intra-string Correlation

In compliance with the model assumption that the inter-string coupling is secondary compared with the intra-string correlation (as shown in Sec. IV, the calculated ratio of the inter- to the intra-string coupling strength is much less than one for the typical glass former glycerol), we shall ignore in a first stage the secondary inter-string interaction $H_2$ to obtain the dynamical behavior of a string using the perturbation theory. Taking into account the linear response theory [41,42] and the Boltzmann principle, the rate equation of $n$ coupled dipoles in an individual $n$-string can be written as (see Appendix I),

$$\frac{d\delta_k}{dt} = -\nu_0 e^{-V_0/T} \sum_{l=1}^{n} M_{kl} \delta_l \qquad (2)$$

where $\delta_k$ is the deviation of the probability from the equilibrium value when the $k$th dipole in the string is at the state $\sigma_k = 1$, and $k, l = 1, \cdots, n$. $M_{12} = M_{nn-1} = 2e^{-2V_1/T} - 1$ and $M_{kk+1} = M_{k+1k} = -1/2$ in the case of $e^{2V_1/T} \gg 1$, a situation that interests us the most in this paper. $M_{kk} = 1$ and the other elements are zero. The factor $\nu_0 e^{-V_0/T}$ is the transition rate between the double-wells and $\nu_0$ is the vibration frequency [18].

Using a unitary transformation, the $n$ coupled equations (Eq.2) are converted into $n$ independent ones that correspond to $n$ individual relaxation modes. Only the mode with the largest relaxation time (called the main mode hereafter) dominates the string relaxation because its relaxation strength is far larger than that of the other modes (called secondary modes), the ratio being about $n$ or larger (see Appendix I). The average string length (~20 to 70 molecules) is much larger than one in the supercooled liquid state (Sec. III-2) and since we are only interested in the $\alpha$-relaxation, we assume that an $n$-string motion can merely be described by the main mode, which is given by,

$$\tau_G(n) = \nu_0^{-1} e^{(V_0 + 2V_1)/T} (n-1)/2 \qquad (3)$$

The effective electric dipole moment associate with this mode is (see Appendix I),

$$\mu(n) = \mu_E R(n)/b \qquad (4)$$

where $\mu_E$ is the contribution of the molecular permanent dipole moment to the effective dipole moment of the main relaxation mode, $R(n)$ is the end-to-end vector amplitude of the $n$-string [36,39], and $b$ is the average distance between molecules in the string.

Eqs.2 and 3 show that the relaxation dynamics of an $n$-string is equivalent to that of an effective dipole, named superdipole (SD) hereafter, whose characteristic relaxation time and electric dipole moment are $\tau_G(n)$ and $\mu(n)$, respectively. An SD has two orientation states, $\sigma = 1$ and $\sigma = -1$ (see Appendix I). Relaxation of the SD involves the visit to $2^n$ orientation states of $n$ dipoles in an $n$-string carried out by hopping across local barriers in the energy landscape [6].

## III-2. String-length Distribution

As shown in Appendix II, the probability, $g_n$, that a dipole is located in an $n$-string is not only determined by the intra-string interaction $H_1$ but also by both the coordination number $z$ and the directional angle of the string $\theta$. For $z = 2$ and $\theta = \pi/2$, the statistic dynamics methods described in Appendix II give the probability $g_n$ for a dipole of the system to belong to an $n$-string as $g_n = n e^{-n/n_0}/n_0^2$, an expression similar to Flory's well-known molecular weight distribution function [43,44]. Notice that $n_0 \equiv e^{V_1/T}/2 >> 1$, provided that the average string length is large enough. Based upon the conditional probability theory, we obtain the same result for the string length distribution $g_n$.

For arbitrary values of $z$ and $\theta$, and also under the condition that the average string length is long enough, $g_n$ becomes the Schulz distribution [43-44] (see Appendix II),

$$g_n = \frac{n^{z_e}}{\Gamma(z_e+1)n_0^{z_e+1}} e^{-n/n_0} \qquad (5)$$

where $z_e = (z-1)\sin\theta$. In this case, the number average of dipoles in the strings is $\bar{n} = \left(\sum_{n=1}^{\infty} g_n/n\right)^{-1} = z_e n_0$ [43-44], where $\bar{n}$ corresponds to the maximum value of $g_n$. Eq.5 becomes the Flory distribution for $z = 2$ and $\theta = \pi/2$.

Shown in Fig.1a are the calculated results for $g_n$, at different temperatures, plotted as a function of $n$ for $z = 7.0$, $V_1 = 640$ K and $\theta = \pi/3.9$, whereas the corresponding $\bar{n}$ *vs* $T$ plot is presented in the inset c of Fig.1a.

### III-3. Inter-string Correlation

As mentioned in Sec. III-1, each string relaxes as an individual SD so that the system can be viewed as an SD gas if the inter-string interaction $H_2$ is ignored. If the secondary $H_2$ and the random distribution of the SDs in space are considered, the system becomes a normal SD liquid. Next, we will use an individual-SD mean-field approach for $H_2$ in consonance with the Debye theory of normal liquids [18]. In other words, the relaxation of the SD liquid is assumed to proceed through SD hopping processes in effective double-wells produced by other SDs. The SD dipole moment $\mu(n)$ (Eq.3) is not altered but the relaxation time of the SD gas state $\tau_G(n)$ (Eq.2) changes to the relaxation time of the SD liquid state $\tau_L(n)$.

According to the SD scenario mentioned in Sec. III-1, the values of $\sigma_k^{mn}$ for all dipoles in a given SD are the same, and we use the symbol $\sigma^{mn}$ to indicate it. Consequently, $H_2$ in Eq.1 can be rewritten as $H_2 = \frac{V_2}{2}\sum_m \sigma^{mn}\sum_{k=1}^{n}\sum_{m'\neq m,l}^{NN(k)} \sigma_l^{m'n'}\cos\alpha_{kl}^{mm'}$. To decrease the inter-string energy corresponding to $H_2$, a given SD will induce local orientational ordering of its surroundings, a process that includes the redistribution of the SDs at $\sigma^{mn} = 1$ and $-1$ states and the change of $\alpha_{kl}^{mm'}$ (Eq.1) corresponding to the variation of the spatial configurations of the strings. On the other hand, the stiffness or rigidity of the strings which prevents such a tendency is described by the persistence length $a_n$, where $a_n = b(C_n+1)/2$ and $C_n$ is the characteristic ratio of the $n$-string [36,39]. Because we are only interested in the SD relaxation, i.e. the redistribution of an SD at $\sigma^{mm}=1$ and $-1$ states, let $p$ be the probability of an SD at the state $\sigma^{mm}=1$. Then $\eta \equiv 2p-1$ is the local order-parameter of the SD liquid.

According to the physical meaning of the persistence length $a_n$ [36,39], the rotations of dipoles in a part of a given $n$-string, shorter than $a_n$, are strongly correlated. So, the $n$ dipoles in an SD can physically be divided into $n*$ sets of dipoles ($n* = nb/a_n$) with $n_\beta$ dipoles in each set ($n_\beta = n/n* = a_n/b$), in such a way that the rotations of different sets are uncorrelated though the rotations of the dipoles in each set are correlated. Thus, $H_2$ can be rewritten as

$$H_2 \approx \frac{V_2}{2}\sum_m \sigma^{mm} n* \sum_{j=1}^{n_\beta} \sum_{m'\neq m,l}^{NN(k)} \sigma_l^{m'n'} \cos\alpha_{jl}^{mm'}$$. Moreover, taking into account the Weiss mean-field theory,

[45] we make the following Weiss effective-internal-field average: $C_\alpha \eta \equiv -\left\langle \sum_{k=1}^{n_\beta} \sum_{m'\neq m,l}^{NN(k)} \sigma_l^{m'n'} \cos\alpha_{kl}^{mm'} \right\rangle_o$,

where $\eta$ and $C_\alpha$ are average quantities related to $\sigma_l^{m'n'}$ and $\cos\alpha_{kl}^{mm'}$, respectively, and $<\cdots>_o$ denotes the average over all the reorientation configurations of the nearest neighbors of the $n_\beta$ dipoles in each set. Then $H_2 \approx -\sum_m n* \overline{V_2} \eta \sigma^{mm}$, where $\overline{V_2} = V_2 C_\alpha$. According to the Weiss mean-field theory [45], the contribution of the inter-SD mean-field $H_2$ to the free energy of an SD is,

$$U_n = -n* \overline{V_2} \eta^2 / 2$$
$$\eta = \tanh\left(\eta \overline{V_2}/T\right) \qquad (6)$$

This equation indicates that $U_n$ diminishes with increasing $n*$, that is, with the decrease of the persistence length $a_n$. Physically, it should be expected that the stiffer the strings are, the more difficult for them is to change their spatial configurations to lower the inter-string energies [36,39], which is consistent with Eq.6. This equation also suggests the existence of a transition temperature $T_{CO}$ and $T_{CO} = \overline{V}_2$. Above $T_{CO}$, $U_n = 0$ indicates that thermal agitation impede any net correlation between the SDs, a situation that corresponds to the SD gas state. Below $T_{CO}$, $U_n < 0$ means that there is a net correlation between the SDs and this behavior corresponds to the SD liquid state. In other words, with decreasing temperature a transition from the SD gas to the SD liquid occurs. We would like to point out that both the string length distribution and the spatial distribution of the strings, all neglected in the mean-field approach discussed above, lead to the strong dispersion of the transition phenomena. Moreover, this kind of inter-SD correlation is a cooperative effect superimposed upon both the zero-order individual-dipole reorientation of the Debye theory and the first-order snakelike motions, so it should be a relatively weak effect. As a result, the mean-field transition temperature and transition phenomenon become, respectively, a crossover temperature $T_{CO}$ and a weak crossover phenomenon. For its reorientation, an SD needs to overcome the inter-SD energy (Eq.6) or effective barrier height of the effective double-well. The energy for an SD to hop between $\sigma = 1$ and $\sigma = -1$ states is equal to $-2U_n$, where the factor 2 arises from the energy increase of both the SD and its surroundings [18].

As mentioned above, the effect of the string length distribution $g_n$ of the SDs on $U_n$ is not considered in the mean-field method. Generally, the relaxation of an SD always corresponds to a dissipation process arising from the distribution fluctuation of the SD at different orientational states caused by thermal agitation, and this time dependent fluctuation leads to variation of the interaction between SDs with time. Specifically, for an SD with short relaxation time, the strong thermal fluctuation of the SD orientational distribution decreases its effective interaction with its surroundings, and vice versa [2,18,41-42]. Thus, the modified factor of the effective activation barrier produced by a $n'$-string on its neighboring $n$-string can be expressed as $\frac{1}{\tau_L(n)} \int_0^{\tau_L(n)} e^{-t/\tau_L(n')} dt = \frac{\tau_L(n')}{\tau_L(n)} \left[ 1 - e^{-\tau_L(n)/\tau_L(n')} \right]$. Taking into account $g_n$, the effective activation barrier height of an SD, $V_E(n)$, and the relaxation time $\tau_L(n)$ of the SD liquid state are given by the following self-consistent equations,

$$\tau_L(n) = \tau_G(n)e^{V_E(n)/T}$$

$$V_E(n) = -2U_n \sum_{n'=1}^{\infty} g_{n'} \frac{\tau_L(n')}{\tau_L(n)}(1 - e^{-\tau_L(n)/\tau_L(n')}) \qquad (7)$$

A calculation method of these equations is given in Appendix III, and the calculated results are shown in Fig.1b and its inset.

## III-4. Model Results

The results of the model show : (1) a polar supercooled liquid is renormalized to a superdipole (SD) normal liquid; (2) the number distribution of the SDs is $h_n = \dfrac{g_n}{n} \Big/ \displaystyle\sum_{n=1}^{\infty} \frac{g_n}{n}$ (see Appendix II and Eq.5), the SD number density in the system being $N_S = N \Big/ \displaystyle\sum_{n=1}^{\infty} nh_n$, where $N$ is the molecular dipole density of the system; and (3) the relaxation time and the effective dipole moment of an SD are $\tau_L(n)$ (Eq.7) and $\mu(n)$ (Eq.4), respectively. By using the same calculation method of the Debye theory [18,41-42], the angular frequency ($\omega$) dependent complex dielectric susceptibility of the system is given by,

$$\chi^*(\omega) = \chi'(\omega) - i\chi''(\omega)$$
$$= \sum_{n=1}^{\infty} \frac{N_S \mu(n)^2}{3T} \frac{h_n}{1 + i\omega\tau_L(n)} = \frac{N\mu_E^2}{3T} \sum_{n=1}^{\infty} \frac{g_n C_n n^{1/5}}{1 + i\omega\tau_L(n)} \qquad (8)$$

## IV. COMPARISON WITH EXPERIMENTS AND DISCUSSION

The results of the model for the α-relaxation of glycerol [11], a typical glass-former polar liquid [3], are shown, in conjunction with the pertinent experimental results [11], in Figs.2 to 3. The parameters used for the model were: $N\mu_E^2 = 3330$ K, $v_0 = 10^{15.4}$ Hz, $V_0 = 2250$ K, $z = 7.0$, $V_1 = 640$ K, $\theta = \pi/3.9$, and $\overline{V}_2 = 297$ K. The model predicts: i) with decreasing temperature, the average relaxation time $\tau_a$ (corresponding to the maximum value of the α-peak) evolves from a high temperature Arrhenius to a low temperature non-Arrhenius (super-Arrhenius) behavior (inset c of Fig.2); ii) the relaxation function crosses over from near exponential to non-exponential (stretched-exponential) response (Figs.2 and 3); and iii) the relaxation strength shows non-Curie features (inset d of Fig.2).

The first characteristic is related to the crossover from the SD gas to the SD liquid. For $T > T_{CO}$, $\tau_a$ is determined by $\tau_G(n)$ (Eqs.3, 6 and 7), which shows Arrhenius behavior (inset c of Fig.2). On the other hand, for $T < T_{CO}$, $\tau_G(n)$ changes to $\tau_L(n)$ due to the net correlation between the SDs (Eqs.6 and 7), which results in a non-Arrhenius or super-Arrhenius behavior. For comparative purposes, the fitting of the Vogel-Fulcher law [13] to experiments is also shown in the inset c of Fig.2, where a clear deviation of the fitting curve from experimental data at high temperatures can be observed [11]. We would like to point out that the crossover mechanism from high temperature Arrhenius to low temperature Super-Arrhenius behavior is not clear [11]. According to Angell et al., the crossover temperature is that one below which the potential energy landscape in the configuration space becomes important [46]. Kim et al. have shown that the crossover temperature could be identified by the first appearance of rotational heterogeneity [47]. The present model presents an alternative interpretation and its relation with the above two pictures needs further study.

For $T > T_{CO}$, $\tau_a$ is also determined by $\tau_G(n)$ (Eq.3), a parameter weakly dependent on the string length $n$ (inset a of Fig.3), so that the relaxation function is characterized by a nearly exponential function (Fig.1b and Fig.2b). However, for $T < T_{CO}$, $\tau_G(n)$, which shows a weak linear dependence on $n$, becomes $\tau_L(n)$. Thus this parameter crosses over from a small $n$ approximate exponential dependence $\tau_L(n) \sim e^{-U_n/T}$ for $n < \overline{n}$ to a large $n$ approximate power law when $n > \overline{n}$ (Figs.1b). Since $\tau_L(\overline{n})$ corresponds to the maximum value of the dielectric loss $\chi''(\omega)$, the small $n$ approximate exponential dependence of $\tau_L(n)$ broadens the high frequency side of $\chi''(\omega)$ more than the large $n$ approximate power law does to the low frequency side of $\chi''(\omega)$ (Eq.8). As a result, the calculated α-peak shows asymmetric features in the frequency domain that render the relaxation function a stretched-exponential in the time domain. Moreover, the small $n$ approximate exponential dependence of $\tau_L(n)$ enlarges with decreasing temperature while the large $n$ approximate power law changes little (Fig.1b), so that the broadening of $\chi''(\omega)$ peak mainly comes from the high frequency side. This conclusion is consistent with the detailed experimental results of Ref.11 shown in Fig.7. Comparisons of our results with the fittings of the empirical Cole-Davidson (CD) law [15] and the Kohlrausch-Williams-Watts (KWW) law [14] to the experimental results of glycerol, at $T = 195$ K, are also shown in Fig.3.

The CD, KWW and the present model fittings clearly deviate from the experiments at the high frequency side of the $\alpha$-peak, and this deviation is known as the excess wing [11,48] (see Figure 3). Although the relaxation strength of the excess wing is about 10 to 100 times smaller than that of the $\alpha$-peak (Fig.3), it is believed that an explanation of such a wing will contribute to the understanding of the $\alpha$-relaxation mechanism. For example, Lunkenheimer et al. have commented in this regard that no commonly accepted explanation for this phenomenon exists, thus remaining one of the great mysteries in the properties of glass-forming materials [11].

As mentioned in the Sec. III-1, an $n$-string in the frame of the present model has $n$ individual relaxation modes. However, we only focus on the main mode that has both the longest relaxation time and largest relaxation strength, compared with the secondary modes, features that lead to the superdipole scenario. The calculations of Appendix II permit to emphasize: (1) the $n$-1 fast relaxation modes omitted in the superdipole scenario appear at the high frequency side of the $\alpha$-peak; (2) the fast modes provide a wider spectrum compared with the $\alpha$-peak because of their relaxation time distribution; and (3) the contributing relaxation strength is about $1/\overline{n}$ times smaller than that of the $\alpha$-relaxation; for example, from the fitting parameters of the relaxation of glycerol it is obtained $\overline{n} = 57$ at T=195K . These results are comparable with the characteristics of the excess wing, and we think that the fast relaxation modes of the strings presumably cause the excess wing. Moreover, with increasing temperature it is expected that the relaxation times associated with the main mode and the secondary modes differ little (see Appendix II), the average string length is shorter and, consequently, the contribution of the fast modes to the spectra becomes important. This interpretation leads to conclude that the excess wing and the $\alpha$-peak gradually overlap forming a single relaxation peak at temperature high enough. Therefore the predictions of the superdipole scenario for the $\alpha$-peak at high temperatures may differ significantly from the experiments, as the data plotted in Figs.2a and 2b show.

The $\alpha$-relaxation strength can be obtained from Eq.8 as $\Delta\varepsilon \equiv \varepsilon_0 - \varepsilon_\infty = \dfrac{N\mu_0^{\,2}}{3T}\displaystyle\sum_{n=1}^{\infty} g_n C_n n^{1/5}$ ,

where $\varepsilon_0$ and $\varepsilon_\infty$ are the permittivities at the low and the high frequency limits, respectively. As shown in the inset d of Fig.2, the increase of the string length with decreasing temperature (inset c of Fig.1a) contributes to the deviation of the relaxation strength from the classical Curie law of the Debye theory. The fittings of the Curie-Weiss-Chamberlin (CWC) law [16] and the Onsager theory to the experiments are also shown in the inset d of Fig.2.

As mentioned above, the present 7-parameter model gives a quantitative description of the experimental results. By comparative purposes, we will discuss the number of parameters involved in fitting experimental data in the temperature-frequency domain by means of some successful empirical laws [13-16]. Fitting the temperature dependence of the average relaxation time (inset c of Fig.2) involves the Vogel-Fulcher law [13] and the Arrhenius relation for which 5 fitting parameters are needed. To fit the relaxation function at different temperatures using the Cole-Davidson [15] or the Kohlrausch-Williams-Watts equations [14], a parameter $\beta$ dependent on temperature is needed, as well at least three temperature independent parameters; among these three latter parameters, one corresponds to the high temperature plateau value and two to the crossover point and the crossover gradient of the $\beta$ value at low temperatures, as shown in Fig.7 of Ref.11. As for the temperature dependence of the relaxation strength, the 2-parameter Currie-Weiss-Chamberlin law [16] does not give a good enough description of the experimental data (inset d of Fig.2) so at least one more parameter is needed to refine the fitting. As a result, the above empirical laws need eleven parameters to fit the experimental data, some of them with unclear physical meaning, four more parameters than the present model. In fact, a self-contained description of the relaxation spectra of the supercooled liquid state in both temperature and frequency domains is equivalent to that of three temperature-dependent quantities, i.e. the average relaxation time and the spectrum width as well as the relaxation strength. The different physical origins of these quantities indicate that we need three sets of temperature independent parameters to describe their complicated temperature dependence, so from a theoretical point of view the seven model parameters of the present scenario looks very reasonable.

As another comparison with our model, let us discuss the number of parameters of the mesoscopic mean-field theory that till now gives the most successful description of the α-relaxation in temperature-frequency domains [24]. In this model, there is a temperature dependent parameter governing the width and the shape of the response, so from a theoretical point of view the theory is not self-contained. If this parameter can be expressed by at least two temperature independent parameters, the total number of parameters in this theory is also seven, the same as in our model.

In what follows, we would like to discuss a little more the parameters of our model: $N\mu_E^2$, $\nu_0$, $V_0$, $z$, $V_1$, $\theta$ and $\overline{V}_2$ (or $V_2$). The first three parameters are the same as those of the Debye theory [18]. During the fitting process, $\nu_0$ is determined from the intersection of the high temperature linear extrapolation of the $\tau_a$ experimental data with the vertical axis (inset c of Fig.2). From a microscopic point of view, $\nu_0$ is the number of times per time unit the thermal agitation of the vibrational modes forces a dipole to overcome the energy barrier. Specifically, the single-dipole process corresponds to the large wave-vector limit of the vibrational modes, homologous to the high frequency limit of the vibrational spectrum, so the corresponding $\nu_0$ should be of such a frequency. The fitting value $\nu_0$ ($=10^{15.4}$ Hz) is in a reasonable error range compared with the scattering experiments [49]. The coordination number $z$ does not appear in the Debye theory [18], and in fact it is a criterion between an individual-particle mean-field theory and the many-body interaction theory, such as the Chamberlin mesoscopic mean-field theory [24]. In decreasing $z$, the width of $g_n$ increases (Eq.5) and consequently the α-peak broadens (Eqs.7 and 8). The fitting value of $z$ ($=7$) is somewhat smaller than that of the random close-packed structure of spheres, but it looks acceptable if the nonspherical characteristic of the glycerol molecules and the influence of the hydrogen bonds are considered [49]. The directional angle of the string, $\theta$, not only determines the effective dipole moment of an SD (Eq.4), but also affects the string length distribution (Eq.5) and the Angell fragility factor $m$ [50] (Eqs.6 and 7). Specifically, with decreasing $\theta$ the average string length and the fragility factor become shorter and smaller, respectively. So, the crossover from fragile to strong glass corresponds to the decrease of $\theta$, i.e. increase of the string stiffness in the frame of our model. Glycerol is a typical glass-former between the fragile and strong limit [3,11,50], so the fitting value of $\theta$ ($=\pi/3.9$) seems to be reasonable.

The fitting values obtained were: $V_0 = 2250$ K = 0.19 eV, $V_1 = 640$ K = 0.055 eV, and $\overline{V}_2 =$

297 K = 0.026 eV. These results indicate that the inter- to intra-string interaction ratio (the topologic anisotropy of the residual-rotational-correlation (RRC) between adjacent dipoles) is $\overline{V}_2/2V_1$=0.23, in agreement with the assumption of our model according to which the inter-string correlation compared with the intra-string interaction is secondary (Sec. II). Moreover, the fact that $V_1/V_0$ =0.24, leads the model to the Debye theory at high temperature. The interactions between molecules of glycerol arise from hydrogen bonding and van der Waals forces, whose values are about 0.25 eV and 0.1 eV, respectively [45,49]. Therefore the fitting values of $V_0$, $V_1$ and $\overline{V}_2$ indicated above lie in acceptable ranges.

Fig.1b shows that the difference between the logarithms of the relaxation times of two adjacent strings of glycerol of lengths 60 and 6 is about 4, at 195K. It is expected that if the molecular dynamics simulation computing time is similar to the relaxation time of the shorter string, the longer string will not relax in the simulation time scale, as mentioned in Sec. II. Moreover, Fig.1a suggests that most molecules belong to long strings ($n > 5$) with large relaxation times, which form slow mobile molecular domains. Therefore, as indicated in Sec. II, the present model does not seem to conflict with the simulations.

From the fitting parameters of glycerol (see caption of Fig.2), the average number of the dipoles in the strings at $T = T_g$ =185K is $\overline{n}$ =69, the latter number reminding the number of structural units intervening in the segmental motions of polymers which is about 20-50 [26,51]. This value corresponds to the end-to-end vector amplitude $\overline{R} \approx \sqrt{C_n}\,\overline{n}^{3/5}b = 30b \sim 9$ nm, a characteristic spatial size of the strings. Of course, $\overline{R}$ will decrease with increasing temperature. Another length scale in the present model is the persistence length of the string $a_n = b(C_n + 1)/2$ arising from the intra-string directional correlation (see Appendix I). In this case $a_n \approx 5.5b \sim 1.6$ nm when the string length is large enough. Furthermore, owing to both the string length distribution and the fluctuation of the strings distribution in space (i.e. some short strings or long strings congregate in space due to the fluctuation), it should be expected that the coupled strings form spatial correlated regions of fractal morphology in the system. Some of them will relax fast and others slow, which prompt us to the well-known concepts of solid-like and liquid-like clusters proposed by Cohen and Grest [20]. It should also be expected that the average spatial size of the regions would be about $\overline{R}$. On the other hand, the well-designed experimental measurements show that the heterogeneous correlation length, i.e. the average spatial size of the clusters, is about 3 to 5 nm for some glass-formers near the glass transition temperature [52], and the theoretical prediction obtained by considering thermal fluctuations within correlated volumes of cooperative regions is about 2 to 7 nm [53]. These results indicate that the model prediction about the average spatial size of the clusters $\overline{R}$ is in an acceptable range.

**ACKNOWLEDGEMENTS**

We thank Dr. P. Lunkenheimer for providing us the experimental data presented here. We also thank W. X. Zhang and Z. Q. Yu for their help as well as Prof. S. H. Li and Dr. W. Li for enlightening discussions. This work was supported by the National Natural Science Foundation of China (Grant No. 10274028).

**APPENDIX I**

First we will give the orientational partition function of an individual $n$-string. For a dipole has two orientational states, $\sigma = 1$ and $\sigma = -1$, the total number of the orientational configurations of the $n$ dipoles in the $n$-string is equal to $2^n$. Let the energy of the $i$th orientational configuration be $E_i$; then the orientational partition function $Q_n$ of the $n$-string is $Q_n = \sum_{i=1}^{2^n} e^{-E_i/T}$. For an individual $n+1$-string, the $2^{n+1}$ total orientational configurations can be built by adding the two "up' and "down" states of a dipole to each end of all the $2^n$ configurations of the $n$-string, so

$$Q_{n+1} = \sum_{i=1}^{2^n} e^{-(E_i+V_1)/T} + \sum_{i=1}^{2^n} e^{-(E_i-V_1)/T} = \left(e^{-V_1/T} + e^{V_1/T}\right)Q_n = \left(e^{-V_1/T} + e^{V_1/T}\right)^n Q_1.$$ For $Q_1 = 2$ we get

$$Q_n = 2\left(e^{-V_1/T} + e^{V_1/T}\right)^{n-1} \tag{I1}$$

Without losing generality and in the linear response regime [41,42], let us consider an $n$-string perturbed by a small enough electric field according to the following history [11,18]

$$F = \begin{cases} F_0, \, t < 0 \\ 0, \quad t \geq 0 \end{cases} \tag{I2}$$

As a representative case of the relaxation equation of an individual $n$-string we will calculate first that of a straight 3-string. We firstly assume the field along the 3-string direction, also keeping the permanent dipole moment along that direction (the general case, forming an angle the permanent dipole moment with the string direction will be discussed in the latter part of this Appendix). Some quantities of the 3-string are shown in Table 1.

Table 1 Configurations, their corresponding energies and normalized existence probabilities of a straight 3-string before and after applying the electric field

| Index | Configurations | $E_j(0)$ | $E_j(F_0)$ | $q_j(t \to -\infty)$ | $q_j(t=0)$ |
|---|---|---|---|---|---|
| 1 | $\rightarrow\rightarrow\rightarrow$ | $-2V_1$ | $-2V_1-3\mu_0F_0$ | $e^{2V_1/T}/Q_3$ | $(1+3\mu_0F_0/T)e^{2V_1/T}/Q_3$ |
| 2 | $\rightarrow\rightarrow\leftarrow$ | $0$ | $-\mu_0F_0$ | $1/Q_3$ | $(1+\mu_0F_0/T)/Q_3$ |
| 3 | $\rightarrow\leftarrow\rightarrow$ | $2V_1$ | $-2V_1-\mu_0F_0$ | $e^{-2V_1/T}/Q_3$ | $(1+\mu_0F_0/T)e^{-2V_1/T}/Q_3$ |
| 4 | $\rightarrow\leftarrow\leftarrow$ | $0$ | $\mu_0F_0$ | $1/Q_3$ | $(1-\mu_0F_0/T)/Q_3$ |
| 5 | $\leftarrow\rightarrow\rightarrow$ | $0$ | $-\mu_0F_0$ | $1/Q_3$ | $(1+\mu_0F_0/T)/Q_3$ |
| 6 | $\leftarrow\leftarrow\rightarrow$ | $2V_1$ | $2V_1+\mu_0F_0$ | $e^{-2V_1/T}/Q_3$ | $(1-\mu_0F_0/T)e^{-2V_1/T}/Q_3$ |
| 7 | $\leftarrow\leftarrow\rightarrow$ | $0$ | $\mu_0F_0$ | $1/Q_3$ | $(1-\mu_0F_0/T)/Q_3$ |
| 8 | $\leftarrow\leftarrow\leftarrow$ | $-2V_1$ | $-2V_1+3\mu_0F_0$ | $e^{2V_1/T}/Q_3$ | $(1-3\mu_0F_0/T)e^{2V_1/T}/Q_3$ |

In Table 1, $E_j(0)$ and $E_j(F_0)$ are, respectively, the energies of the *jth* configurations in absence and in presence of the electric field *F*. The parameter $q_j = e^{-E_j/T}/Q_n$ is the normalized existence probability of the *jth* configuration according to the Boltzmann principle. $q_j(t \to -\infty)$ and $q_j(t=0)$ are the values at time $t \to -\infty$ (without the field *F*) and $t=0$, respectively.

After suddenly switching off the electric field at time $t=0$, $q_j$ will gradually recover from the value of $q_j(t=0)$ to the value of $q_j(t \to -\infty)$ by transformation through different configurations. For a single-dipole hopping process during which only one dipole in the *n*-string changes its orientational state during the transformation from the *ith* to the *jth* configuration, the transfer probability per time unit is equal to $\nu_0 e^{-V_0/T} q_i \dfrac{e^{-E_j}}{e^{-E_i} + e^{-E_j}}$, where $\nu_0 e^{-V_0/T}$ is the jump probability per time unit of a dipole that by effect of the thermal fluctuation gets higher energy than $V_0$ to escape from the well, the term $\dfrac{e^{-E_j}}{e^{-E_i} + e^{-E_j}}$ indicates the redistribution probability of the dipole at the *jth* configuration after it escaped from the well (expression based upon the Boltzmann principle), and $q_i$ means that the larger is the probability of the initial configuration, the higher is the transfer rate to the end configuration. However, for a hopping process of multi-dipoles, e.g. *m* dipoles changing simultaneously their orientational states during the transformation from one configuration to another, the transfer probability per time unit is equal to $\nu_0 \left(e^{-V_0/T}\right)^m q_i$. Since $e^{-V_0/T} \ll 1$ [18] (see also the fitting parameters in Sec. IV), the contribution of this multi-dipole hopping process to the relaxation is negligible in comparison with that of the single-dipole. Based upon detailed mathematical calculations [54], the following rate equations describing the transformation between different configurations of the 3-string are obtained,

$$\frac{d}{dt}\begin{bmatrix} q_1 \\ q_2 \\ q_3 \\ q_4 \\ q_5 \\ q_6 \\ q_7 \\ q_8 \end{bmatrix} = \nu_0 e^{-V_0/T} \begin{bmatrix} -(2u+v) & 1-u & 1-v & 0 & 1-u & 0 & 0 & 0 \\ u & -3/2 & 0 & 1/2 & 0 & 1-u & 0 & 0 \\ v & 0 & -(3-2u-v) & u & 0 & 0 & u & 0 \\ 0 & 1/2 & 1-u & -3/2 & 0 & 0 & 0 & u \\ u & 0 & 0 & 0 & -3/2 & 1-u & 1/2 & 0 \\ 0 & u & 0 & 0 & u & -(3-2u-v) & 0 & v \\ 0 & 0 & 1-u & 0 & 1/2 & 0 & -3/2 & u \\ & & & 1-u & & 1-v & 1-u & -(2u+v) \end{bmatrix} \begin{bmatrix} q_1 \\ q_2 \\ q_3 \\ q_4 \\ q_5 \\ q_6 \\ q_7 \\ q_8 \end{bmatrix} \quad (I3)$$

where $u = 1/\left(1 + e^{2V_1/T}\right)$ and $v = 1/\left(1 + e^{4V_1/T}\right)$. The fitting value of $V_1$ determined by comparing the results of the present model (Sec. IV) with those experimentally obtained for glycerol shows that $e^{2V_1/T} \gg 1$ in the temperature range of interest (from 185 to 400 K), so that $u \approx e^{-2V_1/T}$ and $v \approx 1$.

Let $p_k$ be the probability when the *kth* dipole in an *n*-string is at the state $\sigma_k = 1$ (assumed along the field direction without losing generality), then $p_1 = q_1 + q_2 + q_3 + q_4$, $p_2 = q_1 + q_2 + q_5 + q_6$, $p_3 = q_1 + q_3 + q_5 + q_7$. In this situation, the deviation $\delta_k$ of $p_k$ from its equilibrium value $p_k(-\infty)$ is: $\delta_1 \equiv p_1(t) - p_1(-\infty)$ , $\delta_2 \equiv p_2(t) - p_2(-\infty)$ and $\delta_3 \equiv p_3(t) - p_3(-\infty)$ . From Eq.I3, the rate equations for the deviations $\delta_k$ of the 3-string are,

$$\frac{d}{dt}\begin{bmatrix} \delta_1 \\ \delta_2 \\ \delta_3 \end{bmatrix} = -v_0 e^{-V_0/T} \begin{bmatrix} 1 & 2u-1 & 0 \\ v-1/2 & 1 & v-1/2 \\ 0 & 2u-1 & 1 \end{bmatrix} \begin{bmatrix} \delta_1 \\ \delta_2 \\ \delta_3 \end{bmatrix} \qquad (I4)$$

with the initial values: $\delta_1(0) = \left(3e^{2V_1/T} + e^{-2V_1/T}\right)\mu_0 F_0 / T Q_3$ , $\delta_2(0) = \left(3e^{2V_1/T} + 2 - e^{-2V_1/T}\right)\mu_0 F_0 / T Q_3$ and $\delta_3(0) = \left(3e^{2V_1/T} + e^{-2V_1/T}\right)\mu_0 F_0 / T Q_3$ .

The solution of Eq.I4 is,

$$\begin{bmatrix} \delta_1 \\ \delta_2 \\ \delta_3 \end{bmatrix} = C_1 \begin{bmatrix} 1 \\ \sqrt{\dfrac{1-2v}{1-2u}} \\ 1 \end{bmatrix} e^{-t/\tau_1} + C_2 \begin{bmatrix} 1 \\ 0 \\ -1 \end{bmatrix} e^{-t/\tau_2} + C_3 \begin{bmatrix} 1 \\ -\sqrt{\dfrac{1-2v}{1-2u}} \\ 1 \end{bmatrix} e^{-t/\tau_3} \qquad (I5)$$

where the eigen-relaxation times for $e^{2V_1/T} \gg 1$ are

$$\begin{aligned} \tau_1 &= v_0^{-1} e^{V_0/T} / \left[1 - \sqrt{(1-2u)(1-2v)}\right] \approx v_0^{-1} e^{(V_0+2V_1)/T} \\ \tau_2 &= v_0^{-1} e^{V_0/T} \\ \tau_3 &= v_0^{-1} e^{V_0/T} / \left[1 + \sqrt{(1-2u)(1-2v)}\right] \approx v_0^{-1} e^{V_0/T} / 2 \end{aligned} \qquad (I6)$$

and $C_1 = \dfrac{1}{2}\left[\delta_1(0) + \sqrt{\dfrac{1-2u}{1-2v}}\delta_2(0)\right]$ , $C_2 = 0$ and $C_3 = \dfrac{1}{2}\left[\delta_1(0) - \sqrt{\dfrac{1-2u}{1-2v}}\delta_2(0)\right]$ . For a 3-string, the polarization vector of the $i$th mode, $P_i(t)$ $(i = 1,2,3)$ and the total polarization $P(t)$ are given by

$$\begin{aligned} P_1(t) &\equiv 2\mu_0 C_1\left(1 + \sqrt{\frac{1-2v}{1-2u}} + 1\right)e^{-t/\tau_1} \approx \frac{9\mu_0^2}{T}e^{-t/\tau_1}F_0 \\ P_2(t) &\equiv 2\mu_0 C_2\left(1 + 0 - 1\right)e^{-t/\tau_2} = 0 \\ P_3(t) &\equiv 2\mu_0 C_3\left(1 - \sqrt{\frac{1-2v}{1-2u}} + 1\right)e^{-t/\tau_3} \approx \frac{\mu_0^2}{T}e^{-t/\tau_3}F_0 \\ P(t) &= 2\mu_0\left(\delta_1 + \delta_2 + \delta_3\right) \equiv P_1(t) + P_2(t) + P_3(t) \end{aligned} \qquad (I7)$$

where $\tau_i$ is the relaxation time of the ith mode. By the same token, the two eigen-relaxation times for a 2-string are,

$$\begin{aligned} \tau_1 &= v_0^{-1} e^{V_0/T} / 2u \approx v_0^{-1} e^{(V_0+2V_1)/T} \frac{1}{2} \\ \tau_2 &= v_0^{-1} e^{V_0/T} / 2(1-u) \approx v_0^{-1} e^{V_0/T} \frac{1}{2} \end{aligned} \qquad (I8)$$

and the polarization vectors of both the whole string $P(t)$ and the $ith$ mode $P_i(t)$ ($i = 1, 2$) are

$$P_1(t) \equiv 2\mu_0 C_1 (1+1) e^{-t/\tau_1} \approx \frac{4\mu_0^2}{T} e^{-t/\tau_1} F_0$$

$$P_2(t) \equiv 2\mu_0 C_2 (1-1) e^{-t/\tau_2} = 0 \qquad\qquad\qquad\quad \text{(I9)}$$

$$P(t) = 2\mu_0 (\delta_1 + \delta_2) = P_1(t) + P_2(t)$$

For a 4-string, the four eigen-relaxation times are,

$$\tau_1 = \nu_0^{-1} e^{V_0/T} \left\{ 1 - \left[ \frac{(1/2-\nu)^2 + (1-2\nu)(1-2u) + (1/2-\nu)C_0}{2} \right]^{1/2} \right\}^{-1} \approx \nu_0^{-1} e^{(V_0 + 2V_1)/T} \frac{3}{2}$$

$$\tau_2 = \nu_0^{-1} e^{V_0/T} \left\{ 1 - \left[ \frac{(1/2-\nu)^2 + (1-2\nu)(1-2u) - (1/2-\nu)C_0}{2} \right]^{1/2} \right\}^{-1} \approx \nu_0^{-1} e^{V_0/T} 2(2+\sqrt{3})$$

$$\tau_3 = \nu_0^{-1} e^{V_0/T} \left\{ 1 + \left[ \frac{(1/2-\nu)^2 + (1-2\nu)(1-2u) - (1/2-\nu)C_0}{2} \right]^{1/2} \right\}^{-1} \approx \nu_0^{-1} e^{V_0/T} \frac{2}{5}(2-\sqrt{3})$$

$$\tau_4 = \nu_0^{-1} e^{V_0/T} \left\{ 1 + \left[ \frac{(1/2-\nu)^2 + (1-2\nu)(1-2u) + (1/2-\nu)C_0}{2} \right]^{1/2} \right\}^{-1} \approx \nu_0^{-1} e^{V_0/T} \frac{1}{2}$$

$$\text{(I10)}$$

whereas the total polarization $P(t)$ and the polarization of the four modes are given by

$$P_1(t) \equiv 2\mu_0 C_1 \left( \frac{4(1-2u)}{C_0 + (1-2\nu)} + 1 + 1 + \frac{4(1-2u)}{C_0 + (1-2\nu)} \right) e^{-t/\tau_1} \approx \frac{16\mu_0^2}{T} e^{-t/\tau_1} F_0$$

$$P_2(t) \equiv 2\mu_0 C_2 \left( \frac{-4(1-2u)}{C_0 - (1-2\nu)} - 1 + 1 + \frac{4(1-2u)}{C_0 - (1-2\nu)} \right) e^{-t/\tau_2} = 0$$

$$P_3(t) \equiv 2\mu_0 C_3 \left( \frac{4(1-2u)}{C_0 - (1-2\nu)} - 1 - 1 + \frac{4(1-2u)}{C_0 - (1-2\nu)} \right) e^{-t/\tau_3} \approx \frac{\mu_0^2}{6T} e^{-t/\tau_3} F_0 \qquad \text{(I11)}$$

$$P_4(t) \equiv 2\mu_0 C_4 \left( \frac{-4(1-2u)}{C_0 + (1-2\nu)} + 1 + 1 + \frac{4(1-2u)}{C_0 + (1-2\nu)} \right) e^{-t/\tau_4} = 0$$

$$P(t) = 2\mu_0 (\delta_1 + \delta_2 + \delta_3 + \delta_4) = P_1(t) + P_2(t) + P_3(t) + P_4(t)$$

Notice that the expressions for the relaxation times of Eq.I11 are given in Eq.I10. For an individual dipole in the double-well potential, the relaxation time is,

$$\tau = \nu_0^{-1} e^{V_0/T} \qquad\qquad\qquad\qquad\qquad\qquad\qquad \text{(I12)}$$

whereas the polarization vector is given by,

$$P(t) = \frac{\mu_0^2}{T} e^{-t/\tau} F_0 \qquad\qquad\qquad\qquad\qquad\qquad \text{(I13)}$$

In principle, we can continue the process indicated above to obtain the relaxation equations for an arbitrary $n$-string. However, the preceding results clearly show some general tendencies on the $n$-string. First, the relaxation equation of a given $n$-string is [54],

$$\frac{d}{dt}\begin{bmatrix} \delta_1 \\ \delta_2 \\ \vdots \\ \vdots \\ \vdots \\ \delta_{n-1} \\ \delta_n \end{bmatrix} = -v_0 e^{-V_0/T} \begin{bmatrix} 1 & 2u-1 & 0 & \cdots & 0 & 0 & 0 \\ v-1/2 & 1 & v-1/2 & 0 & \vdots & 0 & 0 \\ 0 & v-1/2 & 1 & \cdots & \vdots & \vdots & 0 \\ \vdots & \vdots & \vdots & \cdots & \vdots & \vdots & \vdots \\ 0 & \vdots & \vdots & \cdots & 1 & v-1/2 & 0 \\ 0 & 0 & \vdots & \cdots & v-1/2 & 1 & v-1/2 \\ 0 & 0 & 0 & \cdots & 0 & 2u-1 & 1 \end{bmatrix} \begin{bmatrix} \delta_1 \\ \delta_2 \\ \vdots \\ \vdots \\ \vdots \\ \delta_{n-1} \\ \delta_n \end{bmatrix} \quad \text{(I14)}$$

Second, the main contribution to the relaxation strength of a given $n$-string comes from the relaxation mode associated with the longest eigen-relaxation time, called main mode hereafter. Compared with this mode, the contributions from the other modes (called secondary modes) are small, about a factor $1/n^2$ or less, as Eqs.I6-I9 shows.

As expected, Eq.I14 indicates that for $V_1/T \to 0$ the $n$ dipoles in an $n$-string are uncorrelated, and all the elements of $M_{kl}$, except $M_{kk} = 1$, are zero. In this case, the original coupled relaxation equations degenerate to $n$ independent equations, each one being similar to that of an individual dipole. On the other hand, for $e^{V_1/T} \to \infty$, the matrix $[M_{kk}]$ in Eq.I14 becomes,

$$\begin{bmatrix} 1 & -1 & 0 & \cdots & 0 & 0 & 0 \\ -1/2 & 1 & -1/2 & 0 & \vdots & 0 & 0 \\ 0 & -1/2 & 1 & \cdots & \vdots & \vdots & 0 \\ \vdots & \vdots & \vdots & \cdots & \vdots & \vdots & \vdots \\ 0 & \vdots & \vdots & \cdots & 1 & -1/2 & 0 \\ 0 & 0 & \vdots & \cdots & -1/2 & 1 & -1/2 \\ 0 & 0 & 0 & \cdots & 0 & -1 & 1 \end{bmatrix}$$

which is the well-known Rouse-Zimm matrix [55]. It can be proved that the determinant of this matrix is zero, indicating that the smallest eigen-value is also zero and the corresponding longest relaxation time is infinite. In this situation, the string will not relax, as intuitively one would expect.

In what follows we will calculate the smallest eigen-value $\lambda/2$ of $[M_{kl}]$ corresponding to the relaxation time of the main mode for $e^{2V_1/T} >> 1$, a case that interests us the most in this paper.

According to the calculated results for 2- to 4-strings (Eqs.I6, I8 and I10), it is expected that $\lambda << 1$ and the corresponding eigen-equation is,

$$|e_{kl}| \equiv \begin{vmatrix} 2-\lambda & 4u-2 & 0 & \cdots & 0 & 0 & 0 \\ -1 & 2-\lambda & -1 & 0 & \vdots & 0 & 0 \\ 0 & 1 & 2-\lambda & \cdots & \vdots & \vdots & 0 \\ \vdots & \vdots & \vdots & \cdots & \vdots & \vdots & \vdots \\ 0 & \vdots & \vdots & \cdots & 2-\lambda & -1 & 0 \\ 0 & 0 & \vdots & \cdots & -1 & 2-\lambda & -1 \\ 0 & 0 & 0 & \cdots & 0 & 4u-2 & 2-\lambda \end{vmatrix} = 0$$

By a set of operations of $(e_{k+1l} - e_{kl})/e_{k1}$, $k = 1 \cdots n$ and $l = 1 \cdots n$, the above equation becomes

$$\begin{vmatrix} 2-\lambda & 4u-2 & 0 & \cdots & 0 & 0 & 0 \\ 0 & 1-3\lambda/2+2u & -1 & 0 & \vdots & 0 & 0 \\ 0 & 0 & 1-5\lambda/2+2u & \cdots & \vdots & \vdots & 0 \\ \vdots & \vdots & \vdots & \cdots & \vdots & \vdots & \vdots \\ 0 & \vdots & \vdots & \cdots & \vdots & -1 & 0 \\ 0 & 0 & \vdots & \cdots & 0 & 1-(2n-3)\lambda/2+2u & -1 \\ 0 & 0 & 0 & \cdots & 0 & 0 & 8u-2(n-1)\lambda \end{vmatrix} = 0$$

From this equation, we get $\lambda = 4u/(n-1)$ and the corresponding relaxation time for the main mode is,

$$\tau = \nu_0^{-1} e^{V_0/T} \frac{2}{\lambda} = \nu_0^{-1} e^{(V_0+2V_1)/T} (n-1)/2 \qquad (I15)$$

The relaxation strength of the main mode obtained from the recurrence relation of Eqs.I7, I9 and I11 for straight strings is $\dfrac{P(0)}{F_0} = \dfrac{(n\mu_0)^2}{T}$. This means that the direction of all dipoles in this mode is the same, i.e. $\sigma_k^{mn} = 1$ for $k = 1 \cdots n$ or $\sigma_k^{mn} = -1$ for $k = 1 \cdots n$, and the effective electric dipole moment $\mu$ of the main mode of such a straight $n$-string is $\mu = n\mu_0$ where $\mu_0$ is the permanent electric dipole moment of each molecule. For an $n$-string distributed in space with end-to-end vector amplitude $R(n)$, the value of $\mu$ can be expressed as [54,56],

$$\mu(n) = \mu_E R(n)/b \qquad (I16)$$

where $\mu_E$ is the contribution of the molecular permanent dipole moment to the effective electric moment of the main relaxation mode, $b$ is the average distance between the dipoles, and according to the Flory-Fisher theory of the self-avoiding free rotational chain, $\left\langle R(n)^2 \right\rangle \approx b^2 C_n n^{6/5}$ for $C_n << n$, where $C_n$ is the so-called characteristic ratio. In this latter expression, the symbol $\left\langle \cdots \right\rangle$ denotes the average over all the spatial configurations of the $n$-string, $C_n = \dfrac{1+\cos\theta}{1-\cos\theta} - \dfrac{2\cos\theta\left(1-\cos^n\theta\right)}{n\left(1-\cos\theta\right)^2}$ for a freely rotating chain, where $\theta$ is the directional angle between consecutive molecules in the strings [39]. Moreover, it should be expected that the change of the end-to-end vector of strings affects less the effective electric moments of the secondary relaxation modes than that of the main relaxation mode, the corresponding ratio between them being approximately $n^{-3/5}$ or even less, as Eqs.I7, I9, I11 and I16 suggest. In other words, the contribution of the main relaxation mode to the relaxation strength is approximately $n$ times larger than that of the secondary relaxation modes for large enough string lengths.

**APPENDIX II**

For illustrative purposes, we discuss two ways to deduce the string length distribution $g_n$ (see the text) for $z = 2$ and $\theta = \pi/2$. The first is exactly based upon statistic dynamics as shown in what follows. Let $m$ be the dipoles number of the system and $h_n^m$ be the probability that the $n$-string exists in the system. As a representative case, diverse configurations for $m$=3 are shown in table 2.

Table 2 Configurations for $m$=3 without considering the orientational states

| Index | Configurations |
|:-----:|:--------------:|
| 1 | —  —  — |
| 2 | —  —…— |
| 3 | —…—  — |
| 4 | —…— …— |

where the symbol "—" expresses a dipole without considering its orientation states, and dot symbols and blank spaces indicate, respectively, interactions and no interactions between dipoles. For the first configuration, the three dipoles are not correlated, so the probability of this configuration is proportional to $Q_1^3$. Moreover, as there are three individual dipoles in this configuration, the contribution to $h_1^3$, $h_2^3$ and $h_3^3$ is proportional to $3Q_1^3$, 0 and 0, respectively. By the same token, the probabilities of the second and third configurations are all proportional to $Q_1Q_2$, and their contributions to $h_1^3$, $h_2^3$ and $h_3^3$ are proportional to $2Q_1Q_2$, $2Q_1Q_2$ and 0, respectively. The probability of the fourth configuration is proportional to $Q_3$, and its contribution to $h_1^3$, $h_2^3$ and $h_3^3$ is proportional to $0$, $0$ and $Q_3$, respectively. So, we obtain $h_1^3 = 2Q_1Q_2 + 3Q_1^3$, $h_2^3 = 2Q_1Q_2$ and $h_3^3 = Q_3$. These results are the same as those deduced from a detailed calculation method [54].

Based upon the same method, the values of $h_n^m$ calculated in terms of the partition function $Q_n$ are shown in table 3 for $m = 1 \rightarrow 6$.

Table 3 Calculated values of $h_n^m$ as function of the partition function $Q_n$ for $m = 1 \rightarrow 6$

| $h_n^m$ | n=1 | n=2 | n=3 | n=4 | n=5 | n=6 |
|---|---|---|---|---|---|---|
| m=1 | $Q_1$ | | | | | |
| m=2 | $2Q_1^2$ | $Q_2$ | | | | |
| m=3 | $2Q_1Q_2 + 3Q_1^3$ | $2Q_1Q_2$ | $Q_3$ | | | |
| m=4 | $6Q_1^2Q_2 + 4Q_1^4 + 2Q_1Q_3$ | $2Q_2^2 + 3Q_1^2Q_2$ | $2Q_1Q_3$ | $Q_4$ | | |
| m=5 | $2Q_1Q_4 + 6Q_1^2Q_3 + 3Q_1Q_2^2$ $+ 12Q_1^3Q_2 + 5Q_1^5$ | $6Q_1Q_2^2 + 4Q_1^3Q_2$ $+ 2Q_2Q_3$ | $2Q_2Q_3 + 3Q_1^2Q_3$ | $2Q_1Q_4$ | $Q_5$ | |
| m=6 | $2Q_1Q_5 + 6Q_1^2Q_4 + 6Q_1Q_2Q_3$ $+ 12Q_1^3Q_3 + 12Q_1^2Q_2^2$ $+ 20Q_1^4Q_2 + 6Q_1^6$ | $2Q_2Q_4 + 6Q_1Q_2Q_3$ $+ 3Q_2^3 + 12Q_1^2Q_2^2$ $+ 5Q_1^4Q_2$ | $6Q_1Q_2Q_3 + 4Q_1^3Q_3$ $+ 2Q_3^2$ | $2Q_2Q_4 +$ $3Q_1^2Q_4$ | $2Q_1Q_5$ | $Q_6$ |

The results of Table 3 lead to the recurrence relation $\dfrac{h_{n+1}^{m+1}}{Q_{n+1}} = \dfrac{h_n^m}{Q_n}$, from which

$$h_n^m = \frac{Q_n}{Q_1} h_1^{m-n+1} \qquad\qquad \text{(II1)}$$

This means that we can calculate $h_n^m$ if we only know $h_1^m$. Also table 3 shows that $h_1^m$ can be written as a sum of the polynomial $h_1^m = \sum_{j=1}^{m} B_j^m$ where the values of $B_j^m$ are given in table 4.

Table 4 Values of $B_j^m$ for the polynomial of $h_1^m$

| $B_j^m / Q_1$ | $j=1$ | $j=2$ | $j=3$ | $j=4$ | $j=5$ | $j=6$ |
|---|---|---|---|---|---|---|
| $m=1$ | 1 | 0 | 0 | 0 | 0 | 0 |
| $m=2$ | 0 | $2Q_1$ | 0 | 0 | 0 | 0 |
| $m=3$ | 0 | $2Q_2$ | $3Q_1^2$ | 0 | 0 | 0 |
| $m=4$ | 0 | $2Q_3$ | $6Q_1Q_2$ | $4Q_1^3$ | 0 | 0 |
| $m=5$ | 0 | $2Q_4$ | $6Q_1Q_3+3Q_2^2$ | $12Q_1^2Q_2$ | $5Q_1^4$ | 0 |
| $m=6$ | 0 | $2Q_5$ | $6Q_2Q_3+6Q_1Q_4$ | $12Q_1Q_2^2+12Q_1^2Q_3$ | $20Q_1^3Q_2$ | $6Q_1^5$ |

From table 4, we obtain

$$B_j^m = \frac{j}{j-1} \sum_{i=1}^{\infty} Q_i B_{j-1}^{m-i} \;,\;\; m \geq j \geq 2 \qquad\qquad \text{(II2)}$$

Then,

$$\begin{aligned} h_1^m &= \sum_{j=1}^{\infty} B_j^m = \sum_{j=2}^{\infty}\sum_{i=1}^{\infty} \frac{j}{j-1} Q_i B_{j-1}^{m-i} \\ &= \sum_{i=1}^{\infty} Q_i \sum_{j=2}^{\infty} \frac{j}{j-1} B_{j-1}^{m-i} \\ &= \sum_{i=1}^{\infty} Q_i \sum_{j=2}^{\infty} \left( B_{j-1}^{m-i} + \frac{1}{j-1} B_{j-1}^{m-i} \right) ,\;\; m \geq 2 \\ &= \sum_{i=1}^{\infty} Q_i \left( h_1^{m-i} + A^{m-i} \right) \end{aligned} \qquad\qquad \text{(II3)}$$

where $A^m \equiv \sum_{j=1}^{m} \frac{1}{j} B_j^m$ for $m \geq 2$. From Eqs.II2-II3 we obtain $A^m = \sum_{i=1}^{m-1} Q_i A^{m-i}$, and

$$A^{m+1} - \frac{Q_2}{2} A^m = \sum_{i=1}^{m} Q_i A^{m-i+1} - \frac{Q_2}{2} \sum_{i=1}^{m-1} Q_i A^{m-i} = Q_1 A^m + \sum_{i=2}^{m} Q_i A^{m-i+1} - \frac{Q_2}{2} \sum_{i=1}^{m-1} Q_i A^{m-i} = Q_1 A^m \quad,\quad \text{These}$$

expressions lead to

$$A^m = (Q_1 + Q_2/2) A^{m-1} = Q_1^2 (Q_1 + Q_2/2)^{m-2} \tag{II4}$$

From Eqs.II3 and II4 we obtain

$$
\begin{aligned}
h_1^m &= \sum_{i=2}^{m-1} Q_i \left( h_1^{m-i} + A^{m-i} \right) + Q_1 \left( h_1^{m-1} + A^{m-1} \right) \\
&= \frac{Q_2}{2} \sum_{i=1}^{m-2} Q_i \left( h_1^{m-1-i} + A^{m-1-i} \right) + Q_1 \left( h_1^{m-1} + A^{m-1} \right) \\
&= \left( Q_1 + \frac{Q_2}{2} \right) h_1^{m-1} + Q_1^3 \left( Q_1 + \frac{Q_2}{2} \right)^{m-3} \quad,\quad m \geq 2 \tag{II5} \\
&= \left( Q_1 + \frac{Q_2}{2} \right)^{m-2} h_1^2 + (m-2) Q_1^3 \left( Q_1 + \frac{Q_2}{2} \right)^{m-3} \\
&= 8 \left( m + \frac{Q_2}{2} \right) \left( 2 + \frac{Q_2}{2} \right)^{m-3}
\end{aligned}
$$

If we assume $h_n \equiv h_n^m \big|_{m \to \infty}$, Eqs.I1 and II5 lead to $\frac{h_{n-1}}{h_n} = 1 + \frac{2}{e^{V_1/T} + e^{-V_1/T}}$. For $e^{V_1/T} >> 1$ and using

the mathematic formula $\lim_{y \to \infty} \left( 1 + \frac{1}{y} \right)^y = e$, we obtain $h_n = c e^{-n/n_0}$, where $n_0 = e^{V_1/T}/2$. The probability that a dipole belongs to an $n$-string in the system is $g_n \sim n h_n$, i.e.

$$g_n = \frac{n}{n_0^2} e^{-n/n_0} \tag{II6}$$

which is the well-known Flory distribution function [43-44].

The second way to calculate $g_n$ for $z=2$ and $\theta = \pi/2$ is based upon the conditional probability theory [57]. Let the probability of two adjacent dipoles forming a 2-string be p, where $p = Q_2 / \left( Q_2 + Q_1^2 \right)$. Then, according to the theory, the probability $h_n$ for $n$ adjacent dipoles forming an $n$-string is $h_n \sim p^n$ and $g_n = c n h_n$, so that we obtain Eq.II6, too. However, this way looks indirect and somewhat unclear.

For arbitrary values of $z$ and $\theta$, the mathematic form of $g_n$ must give the Flory distribution when $z=2$ and $\theta = \pi/2$. One possible form of $g_n$ is the Schulz distribution [43-44]. Moreover, when the average string length is large enough, there exists the topologic termination effect during the string formation. This effect arises from the geometric character of the strings, i.e. topologic quasi-one-dimensional, and only the end of a string can connect with each other. If the average string length is large enough, the coordination dipoles of a string end may all belong to the inner parts of other strings, which lead to the termination of string formation. This kind of topologic termination reminds the termination effect during the polymerization processes, where the chain length distribution is described by the Schulz distribution [43-44]. Based upon the above discussion, the Schulz distribution of the string length seems to be appropriate.

As mentioned above, the string formation is restricted by both the topologic structures and the dynamical conditional probability, which are closely related to two factors, (1) the effective coordination number $z_e \equiv (z-1)\sin\theta$ because the restriction of the directional angle of the rotation chain causes that only part of the $z$ coordination dipoles can form strings with a given dipole, and (2) the intra-string interaction $H_1$. By taking into account the topologic restriction in the course of string formation, it should be expected that the increase of $g_n$ is proportional to both the number of string ends $g_n/n$ and $z_e$, i.e. $g_{n+1} - g_n \sim z_e\, g_n/n$. On the other hand, the probability related to the intra-string interaction $H_1$ is $g_{n+1} - g_n \sim -g_n/n_0$ (it can be obtained from Eq.II6 by deducing the string length distribution for $z=2$ and $\theta = \pi/2$). By colligating these two aspects, it is obtained that $g_{n+1} - g_n = c_1 z_e\, g_n/n - c_2\, g_n/n$, with $g_n \sim n^{c_1 z_e} e^{-c_2 n/n_0}$ under the $n \gg 1$ condition, where $c_1$ and $c_2$ are constants independent of $n$. By recurring this formula to the Flory distribution for $z=2$ and $\theta = \pi/2$, we have $c_1 = 1$ and $c_2 = 1$. Finally we get,

$$g_n = \frac{n^{z_e}}{\Gamma(z_e + 1) n_0^{\,z_e+1}} e^{-n/n_0} \qquad (II7)$$

where $\Gamma(\cdots)$ is the Gamma function, which is just the Schulz distribution. The number average of dipoles in the strings is $\bar{n} = \left( \sum_{n=1}^{\infty} g_n / n \right)^{-1} = z_e n_0$ [43-44], where $\bar{n}$ corresponds to the maximum value of $g_n$. We would like to point out that at high enough temperature, where the string length is very short so that the serializing approximation of $n$ used above is invalidated, $g_n$ deviates from the Schulz distribution and it looks more likely the Flory distribution.

**APPENDIX III**

In principle, Eq.7 can be calculated numerically. However, the numerical calculations will deal

with hundreds of coupled non-linear equations at low temperatures, so the computing time could be

either prohibitively large or at best cumbersome. For example, at $T = T_g = 185$ K, $\bar{n} = 69$ (see the

Sec. IV) for glycerol, a value that corresponds to the maximum value of $g_n$, and the number of

equations is about 700. Moreover, the convergence conditions are very strict due to the nonlinear

dependence of $\tau_L(n)$ on $n$. In fact, we cannot find the numerical solutions of Eq.7 for $\bar{n} > 10$

based upon a standard numerical calculation program. We have proceeded to the use of a variational calculus method to solve this problem.

Taking into account the physical meaning of Eq.7, we propose the mean-field for the sum in

$V_E(n)$ as,

$$\sum_{n'=1}^{\infty} g_{n'} \frac{\tau_L(n')}{\tau_L(n)} \left[ 1 - e^{-\tau_L(n)/\tau_L(n')} \right] \approx \left[ \frac{\bar{\tau}_L}{\tau_L(n)} \right]^{\gamma} \left[ 1 - e^{-[\tau_L(n)/\bar{\tau}_L]^{\gamma}} \right] \qquad (III1)$$

Clearly, $\bar{\tau}_L$ and $\gamma$ correspond, respectively, to the average relaxation time of the SDs and their

distribution, and they can be obtained by variational calculus. Actually, from Eqs.III1 and Eq.7, we

obtain the mean-field relaxation time $\tau_L^m(n)$ of the SDs as

$$\tau_L^m(n) = \tau_G(n) \exp \left\{ -\frac{2U_n}{T} \left[ \frac{\bar{\tau}_L}{\tau_L(n)} \right]^{\gamma} \left[ 1 - e^{-(\tau_l(n)/\bar{\tau}_L)^{\gamma}} \right] \right\} \qquad (III2)$$

Moreover, we get the approximate solution $\tau_L^a(n)$ of $\tau_L(n)$ from Eqs.7 and III as,

$$\tau_L^a(n) = \tau_G(n) \exp \left\{ -\frac{2U_n}{T} \sum_{n'=1}^{\infty} g_{n'} \frac{\tau_L^m(n')}{\tau_L^m(n)} \left[ 1 - e^{-\tau_L^m(n)/\tau_L^m(n')} \right] \right\} \qquad (III3)$$

The standard deviation $E_R$ can be defined as $E_R = \sum\limits_{n=1}^{\infty} \left[ \ln \frac{\tau_L^m(n)}{\tau_L^a(n)} \right]^2$, and the variational calculus gives $\delta E_R = \frac{\partial E_R}{\partial \bar{\tau}_L} \delta \bar{\tau}_L + \frac{\partial E_R}{\partial \gamma} \delta \gamma = 0$. The parameters $\bar{\tau}_L$ and $\gamma$ can be obtained from,

$$\begin{cases} \dfrac{\partial E_R}{\partial \bar{\tau}_L} = 0 \\ \dfrac{\partial E_R}{\partial \gamma} = 0 \end{cases} \tag{III4}$$

Eq.III4 indicates that the best expectation values of $\bar{\tau}_L$ and $\gamma$ correspond to the minimum value of $E_R$.

Taking into account the characteristic of Eq.7 and in order to decrease the calculation errors, we use the following formula to determine $\tau_L(n)$,

$$\tau_L(n) = \left[ \tau_L^m(n) \tau_L^a(n) \right]^{1/2} \tag{III5}$$

Shown in the inset d of Fig.1a are $\tau_L^m(n)$, $\tau_L^a(n)$ and $\tau_L(n)$, respectively. It can be seen that the variational calculus described above gives quite good solutions of the self-consistent Eq.7.

Englewood Cliffs, NJ, 1992; W. Stockmayer, *Pure Appl. Chem.* **15**, 539(1967); E. Riande and R. Díaz-Calleja, Electrical Properties of Polymers, Marcel Dekker, Inc., New York, 2004.

**FIGURE CAPTIONS**

**Fig.1**  Shown in Fig.1a are the theoretical string length distribution $g_n$ (Eq.5) as a function of the string length $n$ at different temperatures $T$ with $z = 7.0$, $V_1 = 640$ K and $\theta = \pi/3.9$. The temperature dependence of the average string length $\bar{n}$ is plotted in the inset (c). A double-logarithmic plot of the theoretical values of $\tau_L(n)$ (Eq.7) *vs* $n$ at several temperatures with $\nu_0 = 10^{15.4}$ Hz, $V_0 = 2250$ K, $z = 7.0$, $V_1 = 640$ K, $\theta = \pi/3.9$, and $\bar{V}_2 = 297$ K are represented in Fig.1b: Shown in the inset (d) are the corresponding relaxation times $\tau_L(n)$, $\tau_L^m(n)$ and $\tau_L^a(n)$ as functions of $n$ at 195 K (see Appendix III).

**Fig.2**  Dielectric constant and loss $\chi''(\omega)$ for the $\alpha$-relaxation of glycerol are shown in the frequency domain, at several temperatures, in Fig.2a and Fig.2b, respectively. The solid curves give the theoretical response obtained from Eq.8 using the following parameters for the model: $N\mu_E{}^2 = 3330$ K, $\nu_0 = 10^{15.4}$ Hz, $V_0 = 2250$ K, $z = 7.0$, $V_1 = 640$ K, $\theta = \pi/3.9$, and $\bar{V}_2 = 297$ K. The circle symbols are experimental results [11]. In the inset (c), symbols represent the experimental average relaxation time $\tau_a$. The dashed, dot and solid lines are fittings to experiments of the Vogel-Fulcher law [13]: $\tau_a = 10^{-14.8} \exp[2309/(T - 129)]$ (second), the Arrhenius relation for high temperatures: $\tau_a = 10^{-15.9} \exp(4700/T)$ (second) and the present model, respectively. In the inset (d), symbols are the experimental reduced relaxation strength $T\Delta\varepsilon = T(\varepsilon_0 - \varepsilon_\infty)$ of the $\alpha$-relaxation, and the dashed, dot and solid lines are fittings to experiments of the Curie-Weiss-Chamberlin (CWC) law [16]:

$T\Delta\varepsilon = 6489T/(T - 100)$, the Onsager theory: $T\Delta\varepsilon = N\mu_0^2(\varepsilon_0 + 2)^2/18$ with $\varepsilon_0 = 4$ and the present model, respectively.

**Fig.3** A double-logarithmic plot of the dielectric loss $\chi''(\omega)$ of the α-relaxation in glycerol at 195 K. The dot, dashed and solid lines are fittings of the Cole-Davidson law [15]: $\chi''(\omega) = \mathrm{Re}\left[65/(1 + i1.2\omega)^{0.58}\right]$, the Fourier transform of the Kohlrausch-Williams-Watts law [14]: $\chi(t) = 76\exp\left[-(1.8t)^{0.62}\right]$, and the present model to experiments, respectively. Shown in the inset (a) are the corresponding relaxation times $\tau_G(n)$ (Eq.3) and $\tau_L(n)$ (Eq.7) as functions of the string length $n$. The corresponding string length distribution $g_n$ (Eq.5) *vs n* is presented in the inset (b).

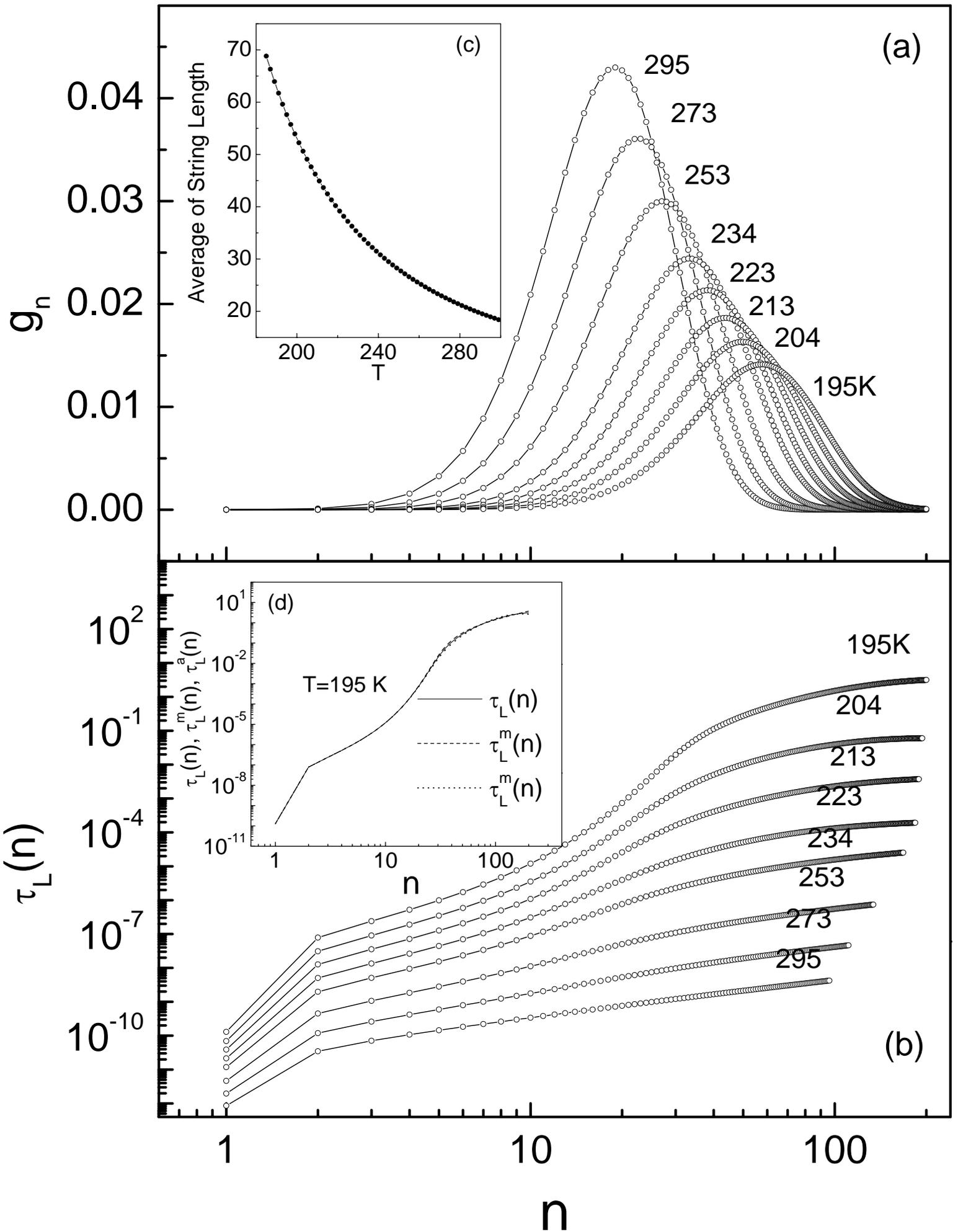

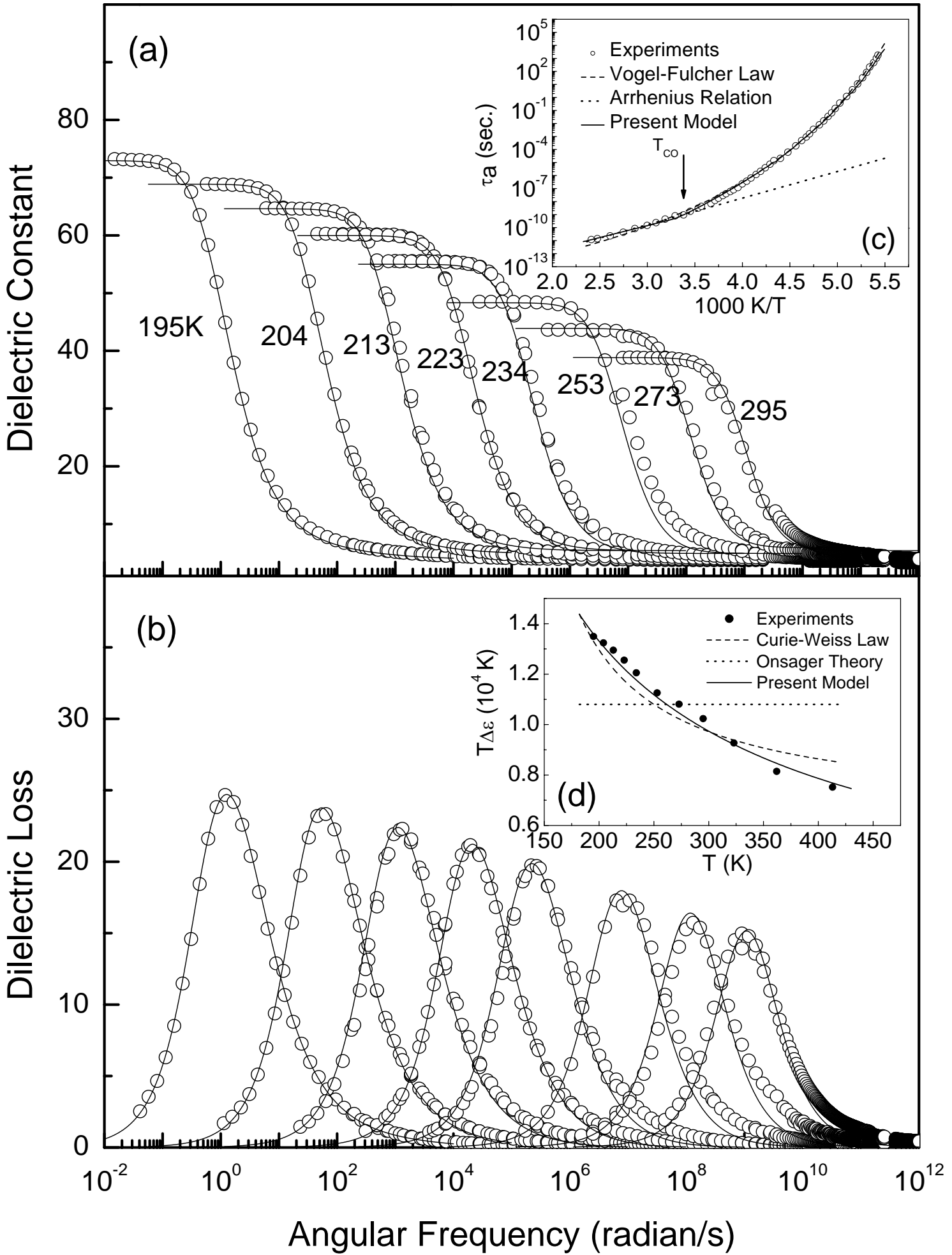

(a)

80

Dielectric Constant

195K    204    213    223    234    253    273    295

(c)

$\tau_a$ (sec.)

○ Experiments
‑ ‑ ‑ Vogel-Fulcher Law
⋯⋯ Arrhenius Relation
── Present Model

$T_{co}$

1000 K/T

(b)

Dilelectric Loss

(d)

$T\Delta\varepsilon$ ($10^4$ K)

● Experiments
‑ ‑ ‑ Curie-Weiss Law
⋯⋯ Onsager Theory
── Present Model

T (K)

Angular Frequency (radian/s)

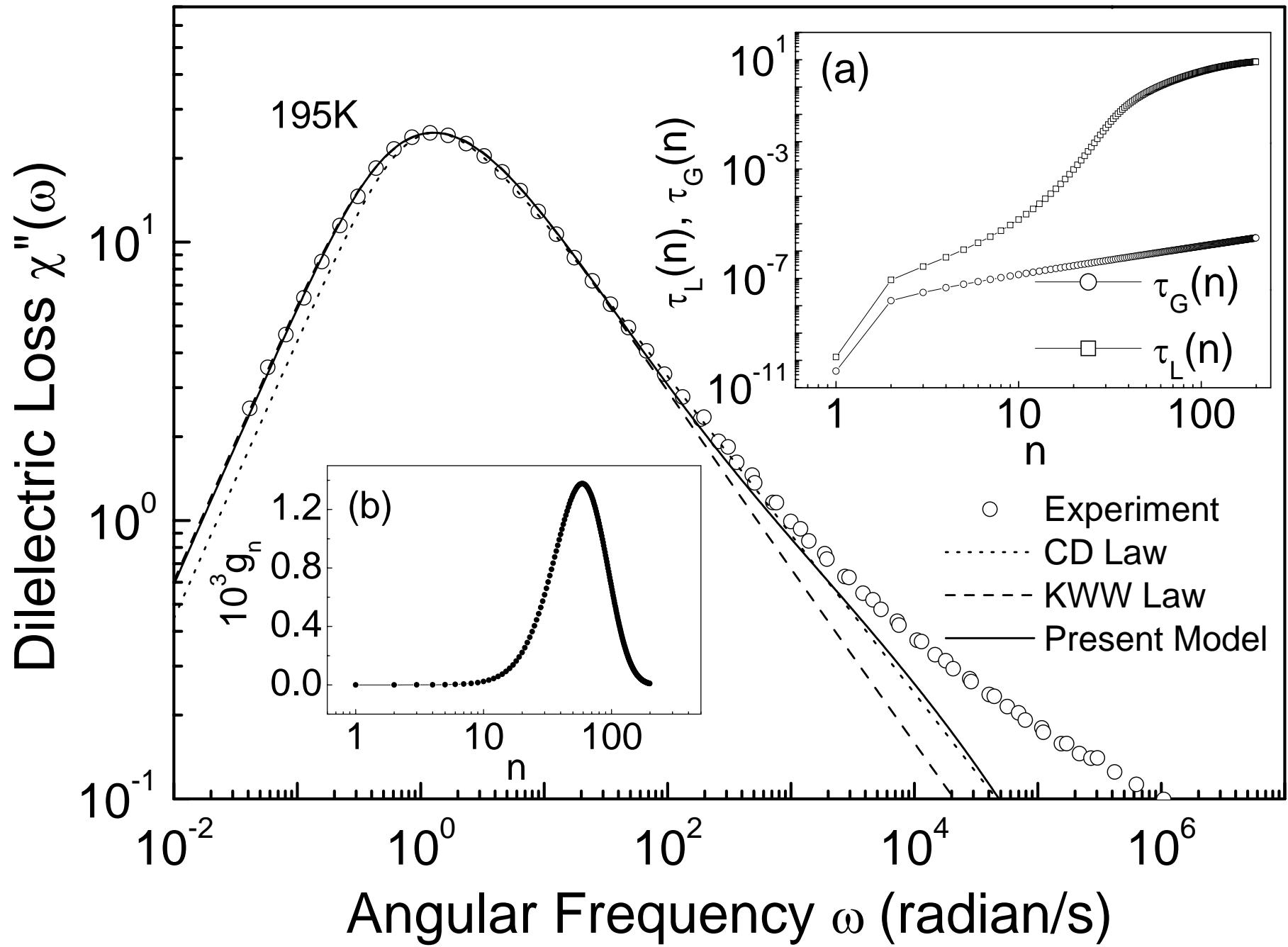